\documentclass[pdflatex,sn-mathphys-num]{sn-jnl}
\usepackage{graphicx}%
\usepackage{multirow}%
\usepackage{amsmath,amssymb,amsfonts}%
\usepackage{amsthm}%
\usepackage{mathrsfs}%
\usepackage[title]{appendix}%
\usepackage{xcolor}%
\usepackage{textcomp}%
\usepackage{manyfoot}%
\usepackage{booktabs}%
\usepackage{algorithm}%
\usepackage{algorithmicx}%
\usepackage{algpseudocode}%
\usepackage{listings}%
\usepackage{lineno}

\theoremstyle{thmstyleone}

\theoremstyle{thmstyletwo}%

\theoremstyle{thmstylethree}%

\begin{document}

\title[]{Quantum Weak Measurement Amplifies Dispersion Signal of Rydberg Atomic System}

\author[1]{\fnm{Yinghang} \sur{Jiang}}
\equalcont{These authors contributed equally.}
\author[1]{\fnm{Jiguo} \sur{Wu}}
\equalcont{These authors contributed equally.}

\author*[2]{\fnm{Meng} \sur{Shi}}\email{shimeng@csu.ac.cn}
\author[1]{\fnm{Hanqing} \sur{Zheng}}
\author[1]{\fnm{Fei} \sur{Guo}}
\author*[1]{\fnm{Zhiguang} \sur{Xiao}}\email{xiaozg@scu.edu.cn}
\author*[1]{\fnm{Zhiyou} \sur{Zhang}}\email{zhangzhiyou@scu.edu.cn}

\affil[1]{\orgdiv{College of Physics}, \orgname{Sichuan University}, \orgaddress{\street{29th Building, Wangjiang Road, Jiuyan Bridge}, \city{Chengdu}, \postcode{610064}, \state{Sichuan}, \country{China}}}

\affil[2]{\orgdiv{Technology and Engineering Center for Space Utilization}, \orgname{Chinese Academy of Sciences}, \orgaddress{\street{9th Building, Dengzhuang Road, Haidian District}, \city{Beijing}, \postcode{100094}, \state{Beijing}, \country{China}}}

\abstract{Rydberg atoms, with their long coherence time and large electric dipole moment, are pivotal in quantum precision measurement. In the process of approaching the standard quantum limit, higher demands are placed on detection schemes. This paper presents a scheme to amplify dispersion signal of Rydberg atomic microwave detection system, using a quantum weak measurement technique together with improved dimensionless pointer. The scheme effectively mitigates the impact of technical noise and can be used to achieve a measurement precision close to the limit set by atomic shot noise in theory. Compared with the superheterodyne method based on transmission detection, our scheme has been experimentally proved to have a sensitivity increase of 5$\sim$6 dB. In this work, the Rydberg dispersion signal amplification mechanism offers a approach to enhance microwave detection sensitivity, which also facilitates deeper investigations into its dynamic processes and further applications of this mechanism in quantum communication and quantum control.}

\maketitle

\section*{Introduction}

Microwave detection plays a crucial role in fields such as radar systems, communication systems, and radio astronomy. As scientific research progresses, there are increasing demands for higher sensitivity in microwave measurements in areas such as particle physics and astrophysics. Traditional metal antennas are limited by electronic thermal noise, and their sensitivity is unable to meet the requirements. Against this backdrop, the Rydberg microwave meter emerges as a promising solution \cite{shi2023near,holloway2014sub,gordon2014millimeter,kubler2010coherent}. Rydberg atoms have large principal quantum numbers and electric dipole moments, making them highly sensitive to external fields \cite{fan2014subwavelength}. Furthermore, their characteristics such as large bandwidth \cite{holloway2017atom,ouyang2023continuous}, traceability \cite{holloway2014broadband}, stealthiness \cite{zhang2018vapor}, etc., make it a promising microwave meter with a great potential. In 2012, Sedlacek et al. first proposed the Rydberg atom microwave measurement technique based on the Electromagnetically Induced Transparency \cite{fleischhauer2005electromagnetically,mohapatra2006coherent} with Autler-Townes (EIT-AT) effect \cite{sedlacek2012microwave,tanasittikosol2011microwave}. This technique utilizes the splitting distance between the two peaks of the Autler-Townes (AT) splitting to measure the strength of the microwave field. Subsequently, more research proposals based on this have been put forward \cite{simons2016using,zhang2018vapor}, such as homodyne detection technique \cite{kumar2017atom}, frequency modulation spectroscopy technique \cite{kumar2017rydberg,hao2023microwave}, three-photon readout scheme \cite{shaffer2018read}. In 2020, Jing et al. proposed the technique of superheterodyne detection \cite{jing2020atomic}, where a known local field and the signal field with a known amplitude and frequency interacted with the atoms. This conversion from measuring microwave fields at the MHz level to measuring optical signals at the kHz level enhances the sensitivity of Rydberg atom microwave meters. In addition to the studies based on the Rydberg system's amplitude, researchers have utilized prism cells \cite{fan2016dispersive} and interferometers \cite{yang2023enhancing} to observe Rydberg atoms dispersion signals \cite{xiao1995measurement}. The direct measurement of Rydberg atoms dispersion signal is of significant importance. In the field of quantum information, the dispersion effect through the process of Electromagnetically Induced Transparency (EIT) enables the realization of slow light \cite{hau1999light} and the storage of photons in atoms \cite{phillips2001storage}. In the laser field, dispersion spectra can be used to achieve higher frequency stability of lasers \cite{purves2004refractive}. In the field of precise measurement, dispersion signals have been demonstrated to play an important role in noise suppression and sensitivity enhancement \cite{berweger2023closed}, and there is a great potential for further advancement in this area.

Quantum weak measurement is a precise measurement technique that has gained significant attention in recent years since its initial conceptualization by Aharonov, Albert, and Vaidman (AAV) in 1988 \cite{aharonov1988result,ritchie1991realization}. It is an advanced technology that offers higher sensitivity for parameter estimation. Weak measurement is characterized by post-selection, and it has found widespread applications in the measurement of various small-scale parameters, such as beam deflection \cite{gorodetski2012weak,dixon2009ultrasensitive}, phase \cite{xu2013phase,brunner2010measuring}, thin film thickness \cite{zhou2012experimental}, velocity \cite{viza2013weak}, frequency \cite{starling2010precision}, temperature \cite{egan2012weak,salazar2015enhancement}, quantum state metrology \cite{PhysRevLett.114.210801,xu2020approaching} and various material parameters \cite{qiu2014determination,zhang2016optical}. Additionally, weak measurement has also been proved to be an effective noise suppression method \cite{lyons2018noise}. This technique is applicable to almost all traditional domains of precise measurement and does not require complex or stringent experimental setups. Both in theory and experiment, it has been demonstrated as a high-precision measurement technique.

 In this paper, we introduce weak measurement techniques based on the Sagnac loop interferometer (SLI) \cite{viza2016complementary,starling2010continuous} in the Rydberg atom system. Traditionally, to observe the dispersion signal of the Rydberg atoms system, Mach-Zehnder interferometer (MZI) is the most popular choice due to its high precision and purely optical structure \cite{song2021enhanced}. However, the combined effects of dispersion and absorption of the probing light in the Rydberg atoms introduce an attenuation factor \cite{yang2023enhancing} in the interference signal determined by the absorption differences of the two arms, which cannot be resolved by any interferometer. An interferometer cannot simultaneously extract phase and amplitude information. However incorporating weak measurement with post selection, we can construct a purely imaginary weak value that eliminates the influence of the absorption differences on the detection results while amplifying the phase difference caused by dispersion. More importantly, using weak value amplification techniques, we can amplify the dispersion signals of Rydberg systems, with an amplification factor inversely proportional to the weak coupling introduced by the SLI . The amplification effect comes from the physical mechanism of quantum weak measurement, rather than from the circuits or noise suppression. Therefore, it is theoretically possible to approximate the atomic shot noise limit by controlling the weak coupling strength. Besides, we propose a method based on this approach that utilizes piezoelectric transducer (PZT) for feedback control, which has been applied in the experiments to suppress the environmental classical noise effectively. After the beam traverses the SLI and passes the post selection, we use an improved dimensionless pointer named intensity-contrast-ratio pointer, to extract the dispersion information. We independently design a Dual-Channel detector which cooperates with the intensity-contrast-ratio circuit to achieve large bandwidth and high precision measurement. In addition, SLI itself has a better stability compared to MZI \cite{mivcuda2014highly}, but to the best of our knowledge, there is no precedent for using it in Rydberg atomic systems and our work is the first to conduct this experiment. To verify the superiority of our measurement scheme, we compare our results with the ones from superheterodyne measurement based on amplitude and dispersion under the same experimental conditions, and our scheme demonstrates an impressive improvement of $5\sim 6$ dB. By optimizing the detector, we can easily achieve higher improvements.

\section*{Results}
\subsection*{Theory}

In previous studies on the dispersion of Rydberg atoms, the technique was primarily based on the MZI \cite{xiao1995measurement, yang2023enhancing}. Although the MZI has the advantages of a simple structure and ease of control, its drawbacks are evident: the significant phase difference caused by only one arm passing through the atomic cell introduces excessive phase noise and mitigates the coherence of the laser, leading to measurement errors in the intensity of the interference light. The use of piezoelectric controlled mirrors to adjust the phase does not effectively eliminate the phase noise. In contrast, the utilization of SLI effectively addresses these issues. SLI allows for the separation of the two arms, both of which enter the atomic cell. By employing quantum weak measurement, the phase difference can be significantly amplified. With PZT feedback control, the method greatly suppresses air disturbances while also suppressing common-mode phase noise, which is not achievable with the MZI \cite{dixon2009ultrasensitive}.

We use two orthogonal linear polarization states $|H\rangle$ and $|V\rangle$ as the system of quantum measurement. Therefore, the observable operator of the system can be represented as: $\textbf{A}=|H\rangle\langle H|-|V\rangle\langle V|$, and the transverse position distribution of the beam $\varphi\left(x\right)=\frac{1}{\sqrt{2 \pi}} w \exp (-\frac{w^2x^2}{4})$ serves as the meter for the measurement, where $ w $ is the waist radius of laser beam and we consider only the x-direction for simplicity. The concept of quantum measurement system and pointer comes from Von Neumann measurement scheme \cite{jordan2024quantum}, and weak measurement adds post selection to it \cite{jordan2015heisenberg,kofman2012nonperturbative}. As shown in Fig. \ref{fig1}a, after the beam enters the SLI consisting of a polarizing beam splitter (PBS) and three mirrors, the clockwise path is the $|V\rangle$ light, while the counterclockwise path represents the $|H\rangle$ light. After completing one round, they converge back to one beam. The crucial point is that the clockwise and counterclockwise paths in the Sagnac loop are not co-propagating. Therefore, in the Rydberg atom system, when we excite one path to the Rydberg state (in this case, the counterclockwise path as mentioned in the article), the influence of the external field will be reflected in the phase difference $\Delta \phi$ between the $|H\rangle$ light and $|V\rangle$ light. The phase information is encoded in the pre-selected state:  $|\psi_{i}\rangle=(e^{i\Delta\phi/2}e^{\Delta\beta}|H\rangle +e^{-i\Delta\phi/2}e^{-\Delta\beta}|V\rangle)/\sqrt{2}$, where $\Delta\beta$ denotes the change in amplitude. So far, we have prepared the pre-selected state, which is essentially the coherent combination of $|H\rangle$ and $|V\rangle$ beams with different amplitudes and phases. It is worth noting that both $\Delta\alpha$ and $\Delta\beta$ are introduced by the Rydberg atomic dynamical process in our experiment. This pre-selected preparation process is also known as biased weak measurement \cite{yin2021improving}. The transverse momentum shift $k$ (along $x$ direction) given to the beam by the mirror introduces entanglement between the system and the meter, which is so-called weak coupling. The evolution matrix can be represented as follows: $\hat{U}=\exp(-ikx\textbf{A})$. The corresponding physical process of the evolution operator is that $|H\rangle$ and $|V\rangle$ light pass through the loop with opposite tiny momentum shifts. The state of the beam exiting the SLI can be expressed as: 
\begin{equation} 
	\begin{aligned}
		\hat{U}|\Psi\rangle=\int d x \varphi\left(x\right)|x\rangle \exp (-i x \mathbf{A} k)|\psi_i\rangle.
	\end{aligned}		
\end{equation}
The post-selected state is represented as: $|\psi_{f}\rangle=(\cos{(\pi/4)}|H\rangle -\sin{(\pi/4)}|V\rangle)/\sqrt{2}$. Post-selection is the core operation in quantum weak measurement, and its essence is a strong projection measurement. It involves projecting different eigenstates of a system onto a specific eigenstate, causing asymmetric interference of the wave functions from the original eignstates, thereby obtaining pointer displacement that are much larger than their eigenvalues.The post-selection projection measurement of a photon system of two polarization states (H and V) can be realized by using a single polarizer in our experiment. The final state wave function of the whole system here can be written as: 
\begin{equation}
	\begin{aligned}
		|\Phi\rangle=\left\langle\psi_f |\hat{U}| \Psi\right\rangle\approx  \left\langle\psi_f \mid \psi_i\right\rangle \int d x \varphi(x)|x\rangle \exp \left(-i x A_w k\right),
	\end{aligned}		
\end{equation}
where a purely imaginary weak value is constructed: $A_w=\frac{\left\langle\psi_f|\mathbf{A}| \psi_i\right\rangle}{\left\langle\psi_f \mid \psi_i\right\rangle}=-i \cot{\Delta\phi/2}$. The expectation value of spatial position can be obtained:
\begin{equation}
	\left\langle x\right\rangle
	=\frac{\left\langle\Phi|x| \Phi\right\rangle}{\left\langle\Phi \mid \Phi\right\rangle}
	=\frac{4 k w^2 e^{i\Delta\phi} \sin{\Delta\phi}}{1-2 e^{2 k^2 w^2+i\Delta\phi}+e^{2 i \Delta\phi}}
	\approx \frac{-\Delta\phi}{k}.
	\label{eq3}
\end{equation}
The conditions for taking an approximation are: $\Delta\phi\ll 1$ and $k^2 w^2\ll 1$. The equation above demonstrates a few interesting results: Firstly, phase information is transferred from the intensity of light to the centroid of light. This is one of the distinguishing features of our measurement scheme compared to traditional interferometers. Secondly, the amplitude parameter $\Delta \beta$ corresponding to the absorption in the atomic cell disappears, which is attributed to the property that weak measurement allows for easy separation of the real and imaginary parts in the measurement. Thirdly, the most interesting result is that weak measurement amplifies the measured phase by a factor of $1/k$. The weak value amplification effect helps us approach the atomic noise limit. 
Atomic shot noise is standard quantum limit of frequency measurement in Rydberg atomic system: $\delta \nu=1/(T\sqrt{N_{at}})$, where $T$ is the time of measurement and $N_{at}$ is the number of atoms participating in the measurement\cite{fan2015atom}. By using the ideal gas equation, it can be obtained that the number of atoms involved in the measurement is around $10^{12}$. In contrast, interferometer exhibits shot noise and radiation pressure noise, and under the conditions of weak probing light, shot noise dominates and is inversely proportional to the square root of the photon number \cite{PhysRevD.23.1693}:  $\delta \phi=1/\sqrt{N_{pho}}$, where $N_{pho}$ is the number of photons participating in the measurement (independent of atom number). The number of photons involved in the reaction can be calculated using the probe laser power and $N_{pho} \approx 10^{15} \mathrm{s^{-1}}$. It is evident that photon shot noise is smaller than atomic shot noise. But the best precision of the experiment has not yet reached the level of atomic shot noise so far, because of the decoherence effect from the environment, the classical noise, the thermal noise in the circuit, etc. Thus, it is difficult to improve the sensitivity of the Rydberg system by only suppressing noise. However, through weak measurement, we can further improve the sensitivity by amplifying the signal. Although the influence of photon vacuum fluctuations cannot be restrained, by controlling the coupling strength to be sufficiently small, the shot noise limit can be achieved in phase measurement.

In order to further illustrate the superiority of our scheme, we compare it with MZI which is based on intensity detection: 
\begin{equation}
	I=a \cos (\Delta \phi)=a \cos \left(\Delta \phi\right),
\end{equation}
where $a$ represents the attenuation caused by the amplitude absorption after passing through the Rydberg system \cite{yang2023enhancing}. Traditional MZI has theoretically highest sensitivity for $\Delta\phi=\pi/2+ n \pi \quad (n=1,2,3...)$, where the intensity and phase exhibit an almost linear relationship. However, for the Rydberg system, the change in amplitude is greater than the change in dispersion (easier to measure), making the attenuation factor $a$ have a significant impact. Furthermore, achieving optimal sensitivity requires perfect alignment of the two arms, which is practically impossible in experiments. As a result, MZI is limited by these shortcomings.

The weak measurement in the SLI, in comparison to the traditional MZI, has the following advantages: Firstly, it amplifies the phase, allowing for a significantly higher sensitivity than the MZI, even without perfect alignment which introduces weak coupling $k$. Secondly, the Sagnac weak measurement scheme does not require the consideration of the absorption difference in the cell between the two arms, as the center of mass is independent of $\Delta \beta$ in theory. Thirdly, SLI introduces a smaller phase difference compared with MZI, which contributes to the preservation of laser coherence. This also provides us with the insight that in interferometer experiments, phase information is not solely contained in the intensity after interference. The information embedded in the intensity distribution may be even more important, and the weak measurement scheme can help to extract it.

\subsection*{Experiment}
\begin{figure}[h]
	\centering
	\includegraphics[width=1.2\textwidth]{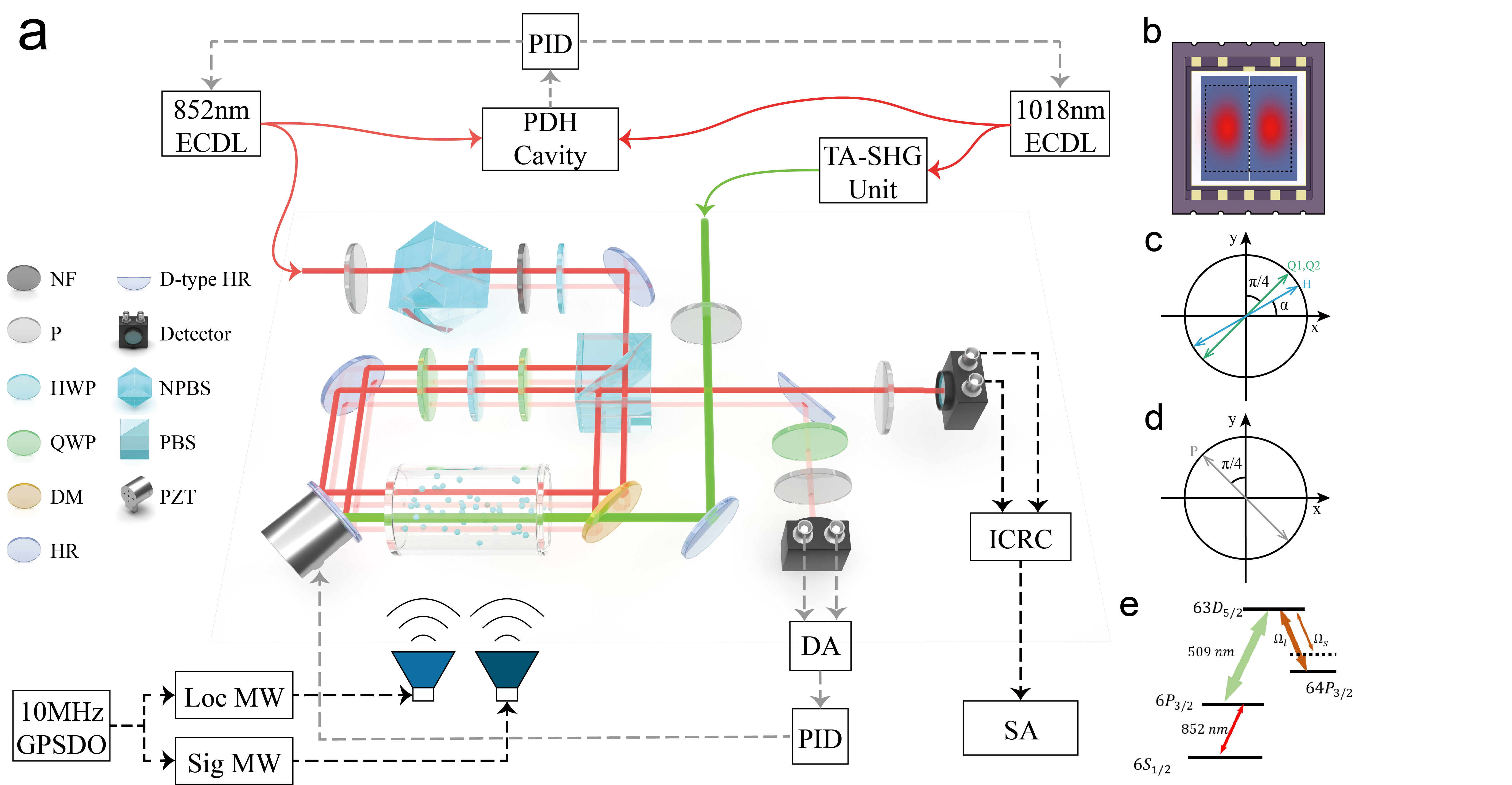}
	\caption{Experimental setup. a. The main optical path structure of the experiment. NF: neutral density filters, P: film polarizer, HWP: half wave plate, QWP: quarter wave plate, DM: dichroic mirrors, HR: dielectric mirror, NPBS: non-polarizing Beamsplitter, PBS: polarizing Beamsplitter, PZT: piezoelectric transducer, PDH Cavity: Pound-Drever-Hall Cavity, ECDL: External cavity and distributed-feedback diode lasers, TA-SHG Unit: tapered amplifier and second-harmonic generation, DA: differential amplifier, PID: Proportional-Integral-Differential controller, ICRC: intensity contrast ratio circuit, SA: spectrum analyzer, GPSDO: GPS disciplined oscillator with Rubidium timebase; b. Frontal sketch of Dual-Channel detector; c. Slow axis angle of the Quarter-Half-Quarter (QHQ) combination,green for quarter wave plate and blue for half wave plate; d. The post-selected angle (gray double arrow); e. Energy level of atoms. State $6S_{1/2}$, $6P_{3/2}$ and $63D_{5/2}$ are coupled by 852 nm and 509 nm laser.The local microwave field ($\Omega_l$) and the weak signal field ($\Omega_s$) are coupled to $64P_{3/2}$.}
	\label{fig1}
\end{figure}

The experimental setup is shown in Fig. \ref{fig1}a. The 852nm probing light is split into two beams, up and down, by a beam splitter (BS). The second polarizer and a half-wave plate together form the pre-selection for weak measurement, resulting in a light linearly polarized at an angle of $45^{\circ}$ with H polarization state. Subsequently, a polarizing beam splitter (PBS) splits the light into clockwise propagating H state light and counterclockwise rotating V state light. Since both beams enter the PBS simultaneously in the entire loop, there are four beams of probing light passing through the Cs cell, exciting Cs atoms from the ground state $6S_{1/2}$ to the excited state $6P_{3/2}$ as shown in Fig. \ref{fig1}b. The quarter-half-quarter wave plate (QHQ) combination in the SLI is used to control the phase difference between H and V light as shown in Fig. \ref{fig1}c (see Method). At this point, the 509 nm coupling light and one of the counterclockwise propagating probing light (the upper beam in the experiment) co-propagate through the Cs cell, coupling Cs atoms to the Rydberg state $63D_{5/2}$, generating EIT, the energy level structure of which is shown in Fig. \ref{fig1}b. After completing one round trip, the two interfering beams recombine and pass through another polarizer forming the post-selection, and are detected by a Dual-Channel detector. One beam that is excited to the Rydberg state serves as the experimental light, while the unexcited beam is reflected by a D-shaped mirror and used as the reference light for feedback control of phase stability. 

To extract the centroid position information, the response time of the traditional image array sensors, such as CCD is far from meeting the requirements. Therefore, we use a custom-designed Dual-Channel detector \cite{Jiang:24} to achieve a high-precision detection. After the post-selection the intensity of the laser beam exhibits a spatial distribution with left and right double peaks as shown in the red part of Fig. \ref{fig1}b. Considering only the $x$-direction, we define the midpoint equidistant to the two peaks as $x=0$, when the two peaks are balanced with the same intensity. The intensity-contrast-ratio is defined as the difference between the light intensity on the left and right sides of the balanced position divided by the sum of the left and right light intensity:
\begin{equation}  
	\begin{aligned}
		\eta=\frac{I_{\leftarrow}-I_{\rightarrow}}{I_{\leftarrow}+I_{\rightarrow}} =\frac{\int_{0}^{\infty}\left\langle\Phi| \Phi\right\rangle dx-\int_{-\infty}^{0}\left\langle\Phi| \Phi\right\rangle dx}{\int_{0}^{\infty}\left\langle\Phi| \Phi\right\rangle dx+\int_{-\infty}^{0}\left\langle\Phi| \Phi\right\rangle dx}=\frac{2 e^{i \Delta\phi}  \operatorname{Erfi}(\sqrt{2} k w) \sin{\Delta\phi}}{1-2 e^{2 k^2 w^2+i \Delta\phi}+e^{2 i\Delta \phi}}
	\end{aligned}
\end{equation}
where $I_{\leftarrow}$ and $I_{\rightarrow}$ represent the total light intensity on the left and right sides, respectively. Under the same approximate conditions as Eq. (\ref{eq3}), we can obtain: $\eta\approx \sqrt{\frac{2}{\pi}}\frac{\Delta\phi}{k w}$. From the results mentioned above, it can be observed that the intensity-contrast-ratio $\eta$ is directly proportional to the phase
difference $\Delta \phi$, and inversely proportional to weak coupling strength $k$ or the laser beam waist radius $w$. The information processing of the dual-channel detector utilizes analog signal processing, which enables a significantly larger bandwidth. Additionally, it also has better detection accuracy, especially when the transverse distribution of the beam is non-uniform. For more details, please refer to the Methods.

\begin{figure}[h]
	\centering
	\includegraphics[width=0.9\textwidth]{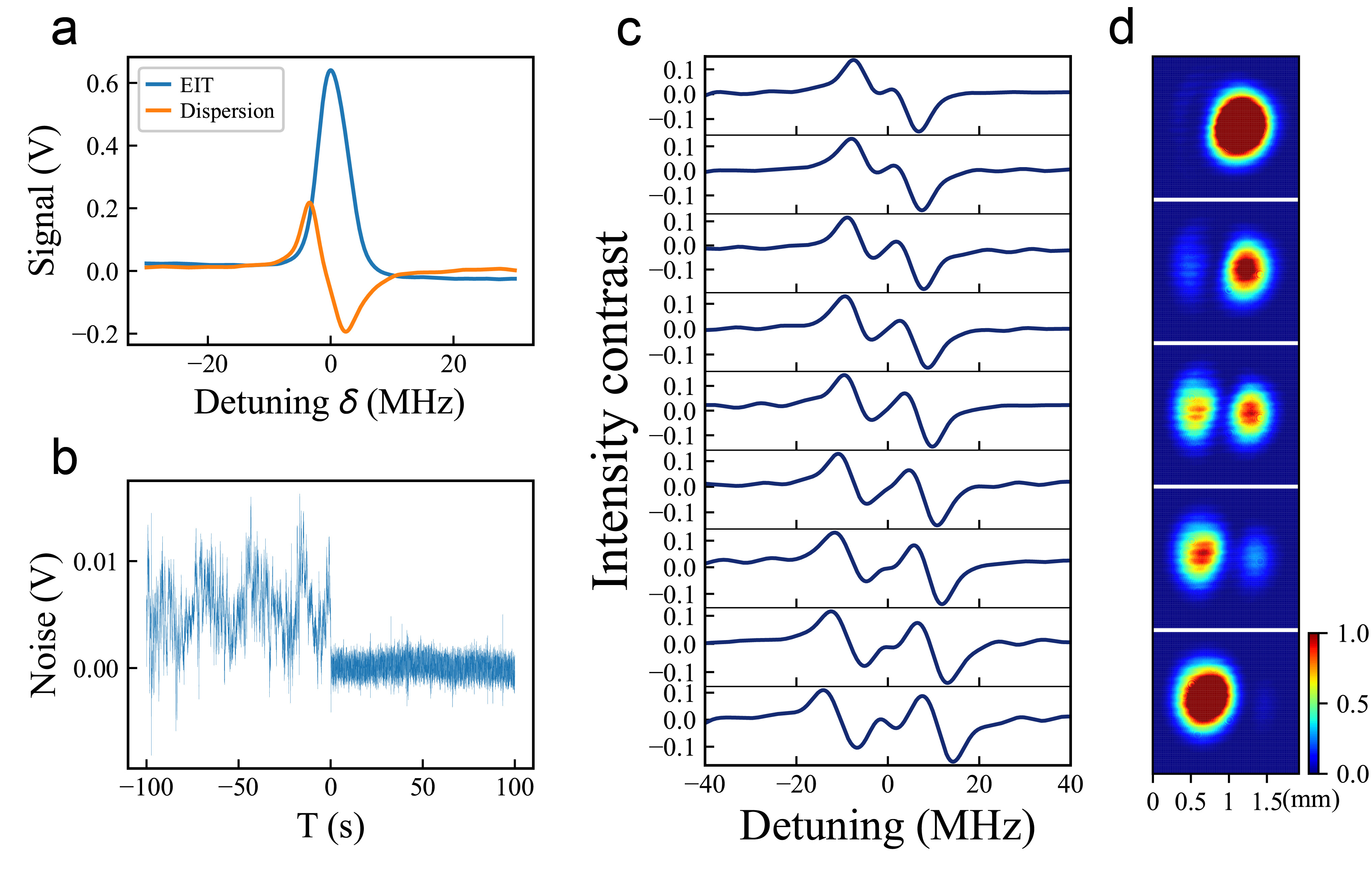}
	\caption{Experimental result. a. The experimentally observed Electromagnetically Induced Transparency (EIT) signal (blue solid line) and the corresponding dispersive signal (orange solid line) are presented; b. The noise signal of the Dual-channel detector. Turned on the piezoelectric at t=0 to feedback control; c. From top to bottom, the change trend of the dispersion signal observed by increasing the amplitude of the applied MW field from $-3\,\mathrm{dBm}$ to $\,\mathrm{dBm}$; d. The change of double-peak-patterned light spot with different detunings of the coupling light. From top to bottom in turn is -100kHz, -50kHz, 0kHz, 50kHz, 100kHz.}
	\label{fig2}
\end{figure}

As shown in Fig. \ref{fig1}a, the tilt angle of the mirror in the lower left corner of the SLI is controlled by PZT. We use a tilted beam splitter (BS) to separate the light into two beams, with the upper beam used for the Rydberg experiment and the lower beam used for feedback control to maintain the stability of SLI. The lower beam passes through the SLI and post-selection, and is adjusted to the balanced state of the double peaks, i.e. $\eta_{\mathrm{con}}=0$, where the subindex ``$\mathrm{con}$" labels the intensity-contrast-ratio for the control beam. After detection by the Dual-Channel detector, the intensity-contrast-ratio signal is used to control the Mirror to align to PZT  through a PID (Proportional-Integral-Derivative) controller to maintain the balanced state. To the best of our knowledge this is the first time that this scheme has been used to perform the optical path stabilization although  previous works \cite{dixon2009ultrasensitive,viza2016complementary,starling2010continuous} have used it to measure the small beam momentum change caused by the mirror. Fig. \ref{fig2}b shows the signal observed in our experiment. We open the PID at $t=0\,\mathrm{s}$ to provide feedback control. Under the condition of keeping the experimental setup unchanged and minimizing external disturbances, it is inevitable that the intensity contrast signal is affected by factors such as instrument fans and building vibrations, leading to significant fluctuations before $t=0$. However, when we activate the feedback control, low-frequency noise is significantly suppressed. The standard deviation of the $\eta_{\mathrm{con}}$ is reduced from $2.9\times 10^{-3}$ to $5.05\times 10^{-4}$ after the suppression and the phase deviations are reduced to less than 0.1 degrees after active stabilization. It is evident that active control of the optical path is necessary in an interferometer.

In the next experiment, we measured the EIT dispersion curve using the SLI weak measurement scheme, as shown in Fig. \ref{fig2}a. Simultaneously, the reference EIT amplitude signal is also displayed. In order to further investigate the dispersion properties of Rydberg systems in AT splitting, we apply a microwave RF signal with a frequency of 2.77 GHz and gradually increase the output power of the RF signal. The changes in dispersion signal are shown in Fig. \ref{fig2}c, from top to bottom, the respective output powers are $-3\,\mathrm{dBm} \sim 5\,\mathrm{dBm}$. The dispersion curves of the two peaks in the AT splitting gradually separate, corresponding to the amplitude signal of the AT splitting. When the power of the  RF signal is weak, the dispersion curve splits but cannot fully separate. However, when the RF amplitude reaches 5 dBm, the dispersion curve completely separates, exhibiting two distinct dispersion curves.  The amplitude and dispersion signals exhibit the expected results predicted perfectly by the KK (Kramers-Kronig) relationship. This result has not been experimentally exhibited in previous research work. Additionally, we lock the probe light frequency at zero detuning point, slowly scan the coupling light from negative detuning (-100 kHz at top) to positive detuning (100 kHz at bottom). The trend of double peaks captured by the CCD camera is shown in Fig. \ref{fig2}d. As we progress from negative to positive detuning,  the double-peak-patterned light spot progressively evolves, starting with a smaller peak on the left and ending with a larger one on the left. The above experimental results perfectly validate the correctness of the theory.

\begin{figure}[h]
	\centering
	\includegraphics[width=1\textwidth]{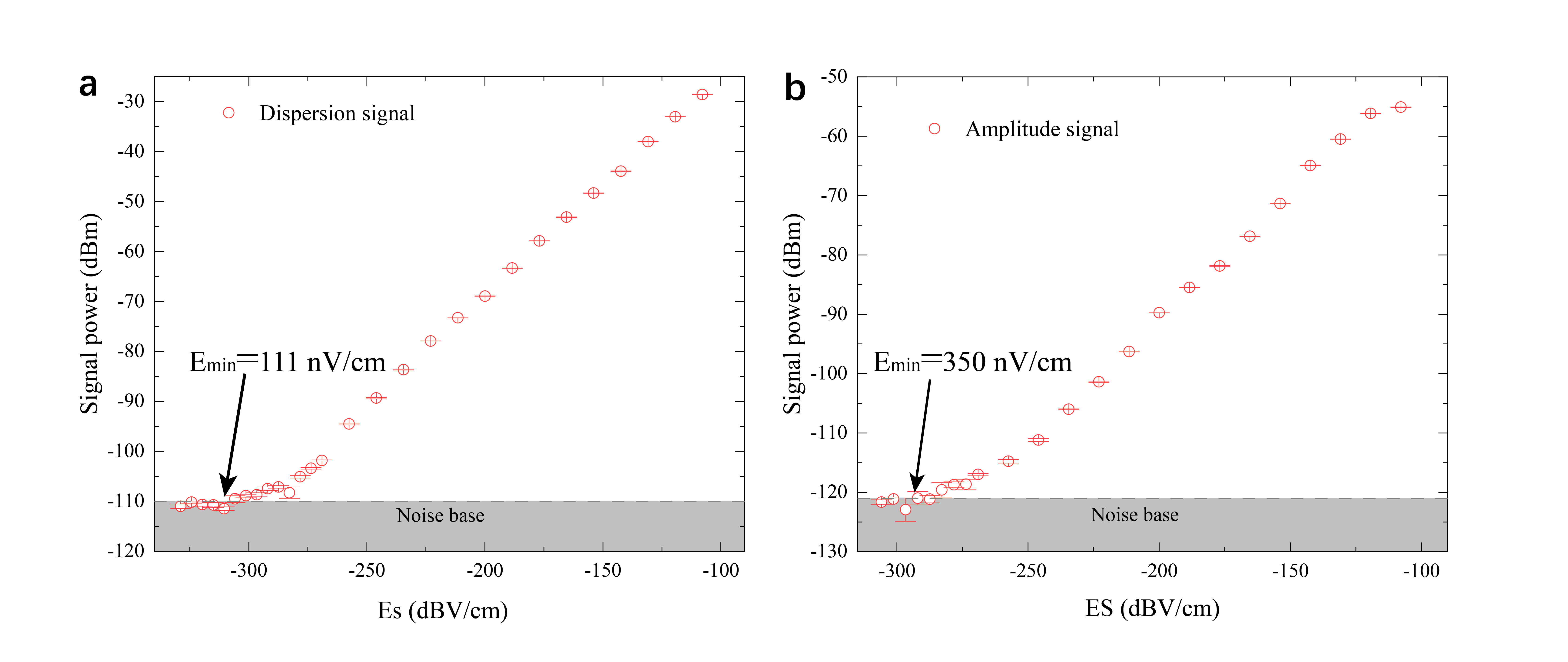}
	\caption{Superheterodyne detection results. Results based on weak measurements SLI (a) and amplitude (b) under the same experimental condition. Red circle is the mean of five measurements, and the space between the two horizontal lines on the red circle is the error,which is the standard deviation of five measurements.  The abscissa is the actual electric field intensity felt by the atom, which is calibrated by the AT splitting (more details can be seen in Method). The ordinate is the received power signal of the spectrometer. As the signal power becomes smaller and smaller, the measurement error gradually increases. Below the dotted line is the noise base, which marks the loss of linearity between signal and electric field amplitude. Above the noise base, the minimum electric field of weak measurement SLI is $111 \, nV/cm$, while amplitude scheme is $350\, nV/cm$. }
	\label{fig3}
\end{figure}

We also compare the weak measurement SLI and amplitude-based superheterodyne detection methods. The superheterodyne method employs optical readout of the beat signal between the signal field and the strong local field. Under the same laboratory noise environment, active frequency stabilization (PDH cavity) and power stabilization (variable optical attenuators), we carried out the experiments with weak measurement SLI and amplitude-based superheterodyne methods, respectively. For detailed experimental procedures, please refer to the Method. Fig. \ref{fig3} presents the experimental results, which demonstrate the superiority of the weak measurement SLI over amplitude-based superheterodyne detection. The minimum detectable field is increased by almost $20 \,\mathrm{dBV \, cm^{-1}}$, the sensitivity of  weak measurement SLI superheterodyne detection is improved by $5\sim 6 \,\mathrm{dB}$. On the one hand, the reason for the improvement of sensitivity is that the dispersion curve has a larger slope maximum in the spectrum than the transmission curve, and on the other hand, it is attributed to the improvement of the signal-to-noise ratio caused by the weak amplification effect. However, it is worth noting that this is far from reaching the limit of weak measurement SLI detection scheme. As we analyze in the theory part, higher sensitivity requires a smaller weak coupling ($k$), which means that the light intensity after post-selection will be weaker. It needs higher requirements for photon noise in the environment and the weak light detection ability of the detector. By using a Dual-Channel detector composed of avalanche photodiodes, we can achieve even greater sensitivity improvements.

\section*{Conclusion}

We have proposed a weak measurement SLI scheme in microwave detection that utilizes quantum weak measurement to amplify the dispersion signal of the Rydberg atomic system. We have discussed the theoretical basis of the scheme and shown that it not only eliminates interference from amplitude absorption but also enhances the dispersion signal through weak coupling amplification mechanisms. The amplification factor is inversely proportional to the weak coupling strength. By elucidating the underlying physical mechanisms, a feasible scheme for approaching the atomic shot noise limit for Rydberg atom systems is proposed. This scheme is shown to be superior to the traditional method using MZI. We also set up the weak measurement SLI experiment to test this scheme. A series of experimental evidence has proven the correctness of the theoretical basis. In addition, we have compared superheterodyne detection based on amplitude with our weak measurement scheme based on dispersion. The results show that our scheme has at least a 6 dB improvement compared with the former. Furthermore, by improving the performance of the detector, one can achieve higher sensitivity. This scheme would be of great interest to a broad range of readers, including researchers in quantum precise measurement, atomic and molecular optics, quantum optics, and other related fields.

\section*{Methods}

\subsection*{Dual-Channel detector}
\begin{figure}[h]
	\centering
	\includegraphics[width=0.7\textwidth]{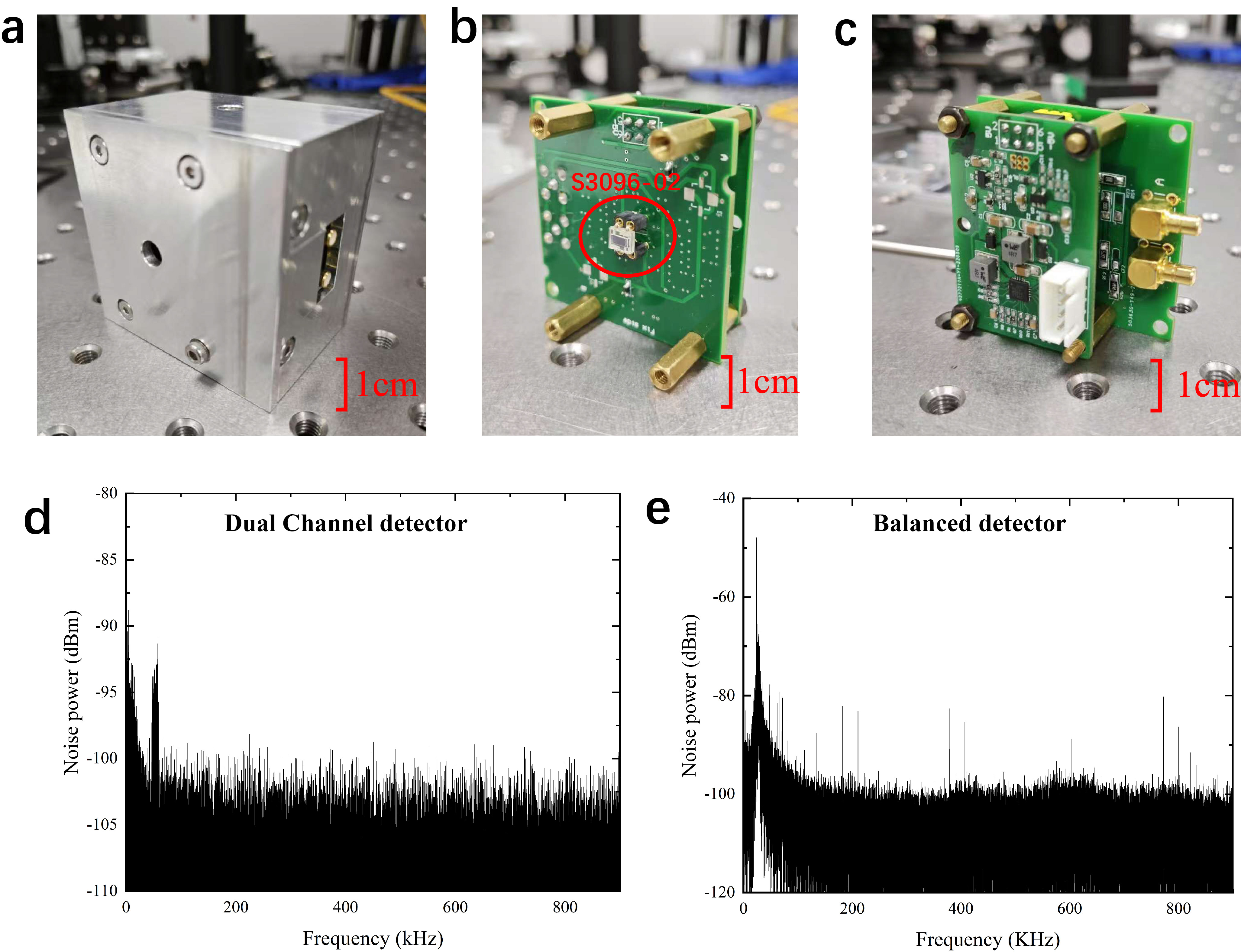}
	\caption{Dual-Channel detector and its noise power spectrum. (a) shielding shell of detector; (b) detector front, S3096-02 in the red circle; (c) detector back; Noise power spectrum of (d) Dual-Channel detector and (e) Balanced detector.   }
	\label{detector}
\end{figure}

The Dual-Channel detector consists of a Si PIN photodiodes S3096-02 from HAMAMATSU and a low noise amplifier circuit. S3096-02 has two compactly arranged $1.2\times3\, \mathrm{mm}$ photosensitive areas, and the element gap is only $30 \,\mu\mathrm{ m}$. Its cutoff frequency is $25\,\mathrm{MHz}$, the maximum dark current is $0.5\,\mathrm{nA}$, and the noise equivalent power is $7.2\times 10^{-15}\,\mathrm{W \, Hz^{-1/2}}$. The amplifier circuit magnifies the signal by 100000 times. The shape of Dual-Channel detector and S3096-02 are shown in Fig.\ref{detector}. We measured the noise power spectrum of the Dual-Channel detector and the balanced detector PDB210A of Thorlabs without signal light input. The background noise of the Dual-Channel detector is slightly higher than that of the balanced detector, because it is more affected by the ambient photon noise. And they all have a low frequency noise of about $30\sim 50 \,\mathrm{KHz}$, which is presumed to be the electrical noise of the laboratory environment.

\subsection*{Feedback control in weak measurement scheme}
The feedback control loop passes through the cell, but the atoms in the path are not excited by coupling light. So the weak coupling process is determined by the deflection angle of the reflector. The description of the measurement system is different from the theoretical part. The evolution operator can be expressed as $e^{ikx\boldsymbol{A}_{F}}$, where $\boldsymbol{A}_{F}=|\circlearrowright\rangle \langle \circlearrowright|-|\circlearrowleft\rangle \langle \circlearrowleft|$, $\{ |\circlearrowright\rangle, |\circlearrowleft\rangle \}$ described beam’s which-path information. The initial state of the system can be described as path state  $|\phi_{Fi}\rangle=(ie^{i\Phi_F/2}|\circlearrowright\rangle+e^{-i\Phi_F/2}|\circlearrowleft\rangle)/\sqrt{2}$, where $\Phi_F$ is the phase difference between $|\circlearrowright\rangle$ and $|\circlearrowleft\rangle$ introduced by QHQ waveplate combination. Through the sagnac loop, the momentum and path state (measurement system) are entangled: $|\Psi_F\rangle=\int dx \varphi(x) |x\rangle \exp(-ik\boldsymbol{A}_\mathrm{F}x) |\phi_{\mathrm{Fi}}\rangle$, where $\varphi(x)$ is the wave function of the meter in the position representation. Postselecting with a final state $|\phi_{\mathrm{Ff}}\rangle=(|\circlearrowright\rangle+i|\circlearrowleft\rangle)/\sqrt{2}$ leaves the state as $\langle \phi_{\mathrm{Ff}}|\Psi_\mathrm{F}\rangle=\left\langle\psi_{\mathrm{Ff}} \mid \psi_{\mathrm{Fi}}\right\rangle \int d x \varphi(x)|x\rangle \exp \left(-i x A_{\mathrm{Fw}} k\right)$, where $A_{\mathrm{Fw}}=-i\cot(\Phi_F) $. The expected value of the beam position is $\langle x\rangle=2kw^2 |A_{\mathrm{Fw}}|$. The above mathematical process is similar to the theoretical part, but the measurement system changes from polarization to which-path information. After postselection, the shift of the beam centroid is amplified by the weak value, which is very sensitive to the change of beam momentum in the sagnac loop. We use this feature in feedback control to adjust the reflector such that the center of mass of the beam is kept constant  constant to maintain the stability of the optical path.

\subsection*{Experiment details}
\begin{figure}[h]
	\centering
	\includegraphics[width=0.6\textwidth]{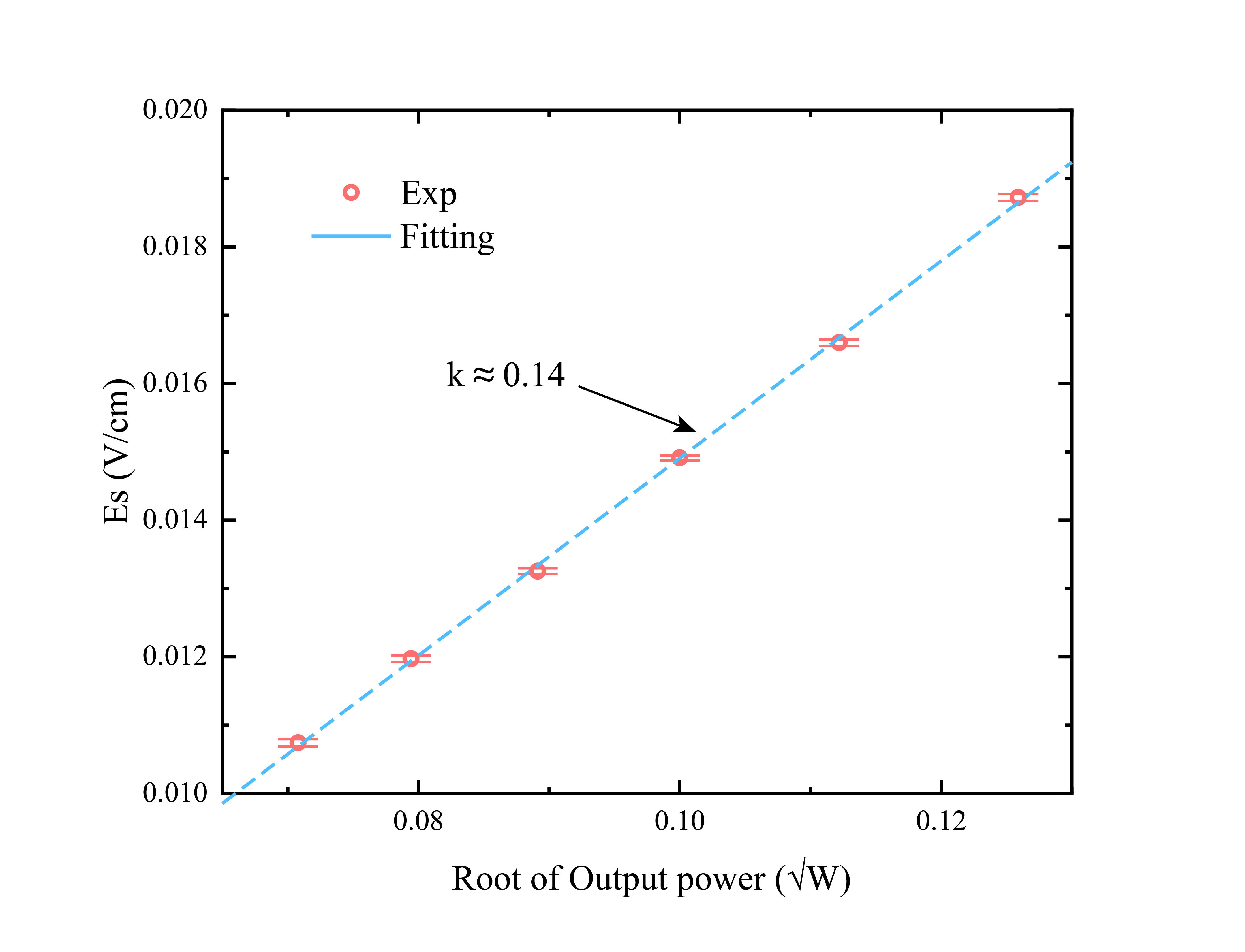}
	\caption{ Calibration curve. The abscissa is the root of input power of the horn, and the ordinate is the electric field intensity calculated by the AT splitting. The red circle is means of experimental measurement results with error bar (standard deviation of three measurements) and blue dotted line is the fitting curve. $k$ is the slope of fitting curve.} 
	\label{calibration}
\end{figure}

In the experiment of microwave sensitivity measurement, the wavelength of the probe light and the coupling light are 852.35 nm and 509.93 nm, respectively. The two lights are all tunable semiconductor lasers with fiber output, and the beam waist is 1.2 mm and 2 mm, respectively, and the frequency is locked by the PDH cavity. The Cs cell is a cylinder with a length of 10 cm. And all the experiments were done at 26 degrees Celsius.  In the amplitude-based superheterodyne detection, we use a variable optical attenuators (VOA) and a PID (SM960) to perform power stabilization, and use a balanced detector to further suppress power noise. We can achieve the best sensitivity by fine tuning the power and frequency of the probe light and coupled light, the output power and frequency of the local microwave field and so on. After optimizing the parameters, the local field frequency is 8.556 GHz which couples Rydberg energy level $ 44D_{5/2}-45P_{3/2} $ , the amplitude is 2.8 dBm, the signal field is 8.566015 GHz, and the frequency difference is 150 kHz. The weak measurement SLI scheme uses one dimensional piezoelectric deflection mirror P33 from ``COREMORROW'' and PID (SM960) to keep the optical path stable. The power of the probe light in front of the cell is 175 $\mu \mathrm{W}$, the power in front of the detector is about 2 $\mu \mathrm{W}$, and the power of the coupled light in front of the cell is 80 $\mathrm{mW}$. The local field frequency is 8.566 GHz and amplitude is 6dBm. The frequency detuning of the probe and coupled light is calibrated by the energy level spacing of $44D_{5/2}$ and $44D_{3/2}$ (839 MHz). The actual electric field strength is determined by AT splitting of Rydberg state $44D_{5/2}$: $E=h f_{\mathrm{AT}}/\boldsymbol{\mu}$, where $h$ is Plank constant, $f_{\mathrm{AT}}$ is AT splitting frequency interval, and $\boldsymbol{\mu}$ is the dipole moment of the Rydberg state. The calibration curve of the experiment as shown in Fig.\ref{calibration}.

\backmatter

\section*{Data availability}
The datasets generated during and/or analysed during the current study are available in the figshare repository\cite{Data}.
\section*{Code availability}
Code available from the corresponding author Z.Z. upon request.

\bigskip


\begin{thebibliography}{54}
	\ifx \bisbn   \undefined \def \bisbn  #1{ISBN #1}\fi
	\ifx \binits  \undefined \def \binits#1{#1}\fi
	\ifx \bauthor  \undefined \def \bauthor#1{#1}\fi
	\ifx \batitle  \undefined \def \batitle#1{#1}\fi
	\ifx \bjtitle  \undefined \def \bjtitle#1{#1}\fi
	\ifx \bvolume  \undefined \def \bvolume#1{\textbf{#1}}\fi
	\ifx \byear  \undefined \def \byear#1{#1}\fi
	\ifx \bissue  \undefined \def \bissue#1{#1}\fi
	\ifx \bfpage  \undefined \def \bfpage#1{#1}\fi
	\ifx \blpage  \undefined \def \blpage #1{#1}\fi
	\ifx \burl  \undefined \def \burl#1{\textsf{#1}}\fi
	\ifx \doiurl  \undefined \def \doiurl#1{\url{https://doi.org/#1}}\fi
	\ifx \betal  \undefined \def \betal{\textit{et al.}}\fi
	\ifx \binstitute  \undefined \def \binstitute#1{#1}\fi
	\ifx \binstitutionaled  \undefined \def \binstitutionaled#1{#1}\fi
	\ifx \bctitle  \undefined \def \bctitle#1{#1}\fi
	\ifx \beditor  \undefined \def \beditor#1{#1}\fi
	\ifx \bpublisher  \undefined \def \bpublisher#1{#1}\fi
	\ifx \bbtitle  \undefined \def \bbtitle#1{#1}\fi
	\ifx \bedition  \undefined \def \bedition#1{#1}\fi
	\ifx \bseriesno  \undefined \def \bseriesno#1{#1}\fi
	\ifx \blocation  \undefined \def \blocation#1{#1}\fi
	\ifx \bsertitle  \undefined \def \bsertitle#1{#1}\fi
	\ifx \bsnm \undefined \def \bsnm#1{#1}\fi
	\ifx \bsuffix \undefined \def \bsuffix#1{#1}\fi
	\ifx \bparticle \undefined \def \bparticle#1{#1}\fi
	\ifx \barticle \undefined \def \barticle#1{#1}\fi
	\bibcommenthead
	\ifx \bconfdate \undefined \def \bconfdate #1{#1}\fi
	\ifx \botherref \undefined \def \botherref #1{#1}\fi
	\ifx \url \undefined \def \url#1{\textsf{#1}}\fi
	\ifx \bchapter \undefined \def \bchapter#1{#1}\fi
	\ifx \bbook \undefined \def \bbook#1{#1}\fi
	\ifx \bcomment \undefined \def \bcomment#1{#1}\fi
	\ifx \oauthor \undefined \def \oauthor#1{#1}\fi
	\ifx \citeauthoryear \undefined \def \citeauthoryear#1{#1}\fi
	\ifx \endbibitem  \undefined \def \endbibitem {}\fi
	\ifx \bconflocation  \undefined \def \bconflocation#1{#1}\fi
	\ifx \arxivurl  \undefined \def \arxivurl#1{\textsf{#1}}\fi
	\csname PreBibitemsHook\endcsname
	
	\bibitem[\protect\citeauthoryear{Shi et~al.}{2023}]{shi2023near}
	\begin{barticle}
		\bauthor{\bsnm{Shi}, \binits{Y.}},
		\bauthor{\bsnm{Ouyang}, \binits{K.}},
		\bauthor{\bsnm{Ren}, \binits{W.}},
		\bauthor{\bsnm{Li}, \binits{W.}},
		\bauthor{\bsnm{Cao}, \binits{M.}},
		\bauthor{\bsnm{Xue}, \binits{Z.}},
		\bauthor{\bsnm{Shi}, \binits{M.}}:
		\batitle{Near-field antenna measurement based on rydberg-atom probe}.
		\bjtitle{Optics Express}
		\bvolume{31}(\bissue{12}),
		\bfpage{18931}--\blpage{18938}
		(\byear{2023})
	\end{barticle}
	\endbibitem
	
	\bibitem[\protect\citeauthoryear{Holloway et~al.}{2014}]{holloway2014sub}
	\begin{botherref}
		\oauthor{\bsnm{Holloway}, \binits{C.L.}},
		\oauthor{\bsnm{Gordon}, \binits{J.A.}},
		\oauthor{\bsnm{Schwarzkopf}, \binits{A.}},
		\oauthor{\bsnm{Anderson}, \binits{D.A.}},
		\oauthor{\bsnm{Miller}, \binits{S.A.}},
		\oauthor{\bsnm{Thaicharoen}, \binits{N.}},
		\oauthor{\bsnm{Raithel}, \binits{G.}}:
		Sub-wavelength imaging and field mapping via electromagnetically induced
		transparency and autler-townes splitting in rydberg atoms.
		Applied Physics Letters
		\textbf{104}(24)
		(2014)
	\end{botherref}
	\endbibitem
	
	\bibitem[\protect\citeauthoryear{Gordon et~al.}{2014}]{gordon2014millimeter}
	\begin{botherref}
		\oauthor{\bsnm{Gordon}, \binits{J.A.}},
		\oauthor{\bsnm{Holloway}, \binits{C.L.}},
		\oauthor{\bsnm{Schwarzkopf}, \binits{A.}},
		\oauthor{\bsnm{Anderson}, \binits{D.A.}},
		\oauthor{\bsnm{Miller}, \binits{S.}},
		\oauthor{\bsnm{Thaicharoen}, \binits{N.}},
		\oauthor{\bsnm{Raithel}, \binits{G.}}:
		Millimeter wave detection via autler-townes splitting in rubidium rydberg
		atoms.
		Applied Physics Letters
		\textbf{105}(2)
		(2014)
	\end{botherref}
	\endbibitem
	
	\bibitem[\protect\citeauthoryear{K{\"u}bler et~al.}{2010}]{kubler2010coherent}
	\begin{barticle}
		\bauthor{\bsnm{K{\"u}bler}, \binits{H.}},
		\bauthor{\bsnm{Shaffer}, \binits{J.}},
		\bauthor{\bsnm{Baluktsian}, \binits{T.}},
		\bauthor{\bsnm{L{\"o}w}, \binits{R.}},
		\bauthor{\bsnm{Pfau}, \binits{T.}}:
		\batitle{Coherent excitation of rydberg atoms in micrometre-sized atomic vapour
			cells}.
		\bjtitle{Nature Photonics}
		\bvolume{4}(\bissue{2}),
		\bfpage{112}--\blpage{116}
		(\byear{2010})
	\end{barticle}
	\endbibitem
	
	\bibitem[\protect\citeauthoryear{Fan et~al.}{2014}]{fan2014subwavelength}
	\begin{barticle}
		\bauthor{\bsnm{Fan}, \binits{H.}},
		\bauthor{\bsnm{Kumar}, \binits{S.}},
		\bauthor{\bsnm{Daschner}, \binits{R.}},
		\bauthor{\bsnm{K{\"u}bler}, \binits{H.}},
		\bauthor{\bsnm{Shaffer}, \binits{J.}}:
		\batitle{Subwavelength microwave electric-field imaging using rydberg atoms
			inside atomic vapor cells}.
		\bjtitle{Optics Letters}
		\bvolume{39}(\bissue{10}),
		\bfpage{3030}--\blpage{3033}
		(\byear{2014})
	\end{barticle}
	\endbibitem
	
	\bibitem[\protect\citeauthoryear{Holloway et~al.}{2017}]{holloway2017atom}
	\begin{barticle}
		\bauthor{\bsnm{Holloway}, \binits{C.L.}},
		\bauthor{\bsnm{Simons}, \binits{M.T.}},
		\bauthor{\bsnm{Gordon}, \binits{J.A.}},
		\bauthor{\bsnm{Wilson}, \binits{P.F.}},
		\bauthor{\bsnm{Cooke}, \binits{C.M.}},
		\bauthor{\bsnm{Anderson}, \binits{D.A.}},
		\bauthor{\bsnm{Raithel}, \binits{G.}}:
		\batitle{Atom-based rf electric field metrology: from self-calibrated
			measurements to subwavelength and near-field imaging}.
		\bjtitle{IEEE Transactions on Electromagnetic Compatibility}
		\bvolume{59}(\bissue{2}),
		\bfpage{717}--\blpage{728}
		(\byear{2017})
	\end{barticle}
	\endbibitem
	
	\bibitem[\protect\citeauthoryear{Ouyang et~al.}{2023}]{ouyang2023continuous}
	\begin{botherref}
		\oauthor{\bsnm{Ouyang}, \binits{K.}},
		\oauthor{\bsnm{Shi}, \binits{Y.}},
		\oauthor{\bsnm{Lei}, \binits{M.}},
		\oauthor{\bsnm{Shi}, \binits{M.}}:
		Continuous broadband microwave electric field measurement in rydberg atoms
		based on the dc stark effect.
		Applied Physics Letters
		\textbf{123}(26)
		(2023)
	\end{botherref}
	\endbibitem
	
	\bibitem[\protect\citeauthoryear{Holloway et~al.}{2014}]{holloway2014broadband}
	\begin{barticle}
		\bauthor{\bsnm{Holloway}, \binits{C.L.}},
		\bauthor{\bsnm{Gordon}, \binits{J.A.}},
		\bauthor{\bsnm{Jefferts}, \binits{S.}},
		\bauthor{\bsnm{Schwarzkopf}, \binits{A.}},
		\bauthor{\bsnm{Anderson}, \binits{D.A.}},
		\bauthor{\bsnm{Miller}, \binits{S.A.}},
		\bauthor{\bsnm{Thaicharoen}, \binits{N.}},
		\bauthor{\bsnm{Raithel}, \binits{G.}}:
		\batitle{Broadband rydberg atom-based electric-field probe for si-traceable,
			self-calibrated measurements}.
		\bjtitle{IEEE Transactions on Antennas and Propagation}
		\bvolume{62}(\bissue{12}),
		\bfpage{6169}--\blpage{6182}
		(\byear{2014})
	\end{barticle}
	\endbibitem
	
	\bibitem[\protect\citeauthoryear{Zhang et~al.}{2018}]{zhang2018vapor}
	\begin{barticle}
		\bauthor{\bsnm{Zhang}, \binits{L.}},
		\bauthor{\bsnm{Liu}, \binits{J.}},
		\bauthor{\bsnm{Jia}, \binits{Y.}},
		\bauthor{\bsnm{Zhang}, \binits{H.}},
		\bauthor{\bsnm{Song}, \binits{Z.}},
		\bauthor{\bsnm{Jia}, \binits{S.}}:
		\batitle{Vapor cell geometry effect on rydberg atom-based microwave electric
			field measurement}.
		\bjtitle{Chinese Physics B}
		\bvolume{27}(\bissue{3}),
		\bfpage{033201}
		(\byear{2018})
	\end{barticle}
	\endbibitem
	
	\bibitem[\protect\citeauthoryear{Fleischhauer
		et~al.}{2005}]{fleischhauer2005electromagnetically}
	\begin{barticle}
		\bauthor{\bsnm{Fleischhauer}, \binits{M.}},
		\bauthor{\bsnm{Imamoglu}, \binits{A.}},
		\bauthor{\bsnm{Marangos}, \binits{J.P.}}:
		\batitle{Electromagnetically induced transparency: Optics in coherent media}.
		\bjtitle{Reviews of modern physics}
		\bvolume{77}(\bissue{2}),
		\bfpage{633}
		(\byear{2005})
	\end{barticle}
	\endbibitem
	
	\bibitem[\protect\citeauthoryear{Mohapatra
		et~al.}{2006}]{mohapatra2006coherent}
	\begin{botherref}
		\oauthor{\bsnm{Mohapatra}, \binits{A.}},
		\oauthor{\bsnm{Jackson}, \binits{T.}},
		\oauthor{\bsnm{Adams}, \binits{C.}}:
		Coherent optical detection of highly excited rydberg states using
		electromagnetically induced transparency.
		arXiv preprint quant-ph/0612200
		(2006)
	\end{botherref}
	\endbibitem
	
	\bibitem[\protect\citeauthoryear{Sedlacek et~al.}{2012}]{sedlacek2012microwave}
	\begin{barticle}
		\bauthor{\bsnm{Sedlacek}, \binits{J.A.}},
		\bauthor{\bsnm{Schwettmann}, \binits{A.}},
		\bauthor{\bsnm{K{\"u}bler}, \binits{H.}},
		\bauthor{\bsnm{L{\"o}w}, \binits{R.}},
		\bauthor{\bsnm{Pfau}, \binits{T.}},
		\bauthor{\bsnm{Shaffer}, \binits{J.P.}}:
		\batitle{Microwave electrometry with rydberg atoms in a vapour cell using
			bright atomic resonances}.
		\bjtitle{Nature physics}
		\bvolume{8}(\bissue{11}),
		\bfpage{819}--\blpage{824}
		(\byear{2012})
	\end{barticle}
	\endbibitem
	
	\bibitem[\protect\citeauthoryear{Tanasittikosol
		et~al.}{2011}]{tanasittikosol2011microwave}
	\begin{barticle}
		\bauthor{\bsnm{Tanasittikosol}, \binits{M.}},
		\bauthor{\bsnm{Pritchard}, \binits{J.}},
		\bauthor{\bsnm{Maxwell}, \binits{D.}},
		\bauthor{\bsnm{Gauguet}, \binits{A.}},
		\bauthor{\bsnm{Weatherill}, \binits{K.}},
		\bauthor{\bsnm{Potvliege}, \binits{R.}},
		\bauthor{\bsnm{Adams}, \binits{C.}}:
		\batitle{Microwave dressing of rydberg dark states}.
		\bjtitle{Journal of Physics B: Atomic, Molecular and Optical Physics}
		\bvolume{44}(\bissue{18}),
		\bfpage{184020}
		(\byear{2011})
	\end{barticle}
	\endbibitem
	
	\bibitem[\protect\citeauthoryear{Simons et~al.}{2016}]{simons2016using}
	\begin{botherref}
		\oauthor{\bsnm{Simons}, \binits{M.T.}},
		\oauthor{\bsnm{Gordon}, \binits{J.A.}},
		\oauthor{\bsnm{Holloway}, \binits{C.L.}},
		\oauthor{\bsnm{Anderson}, \binits{D.A.}},
		\oauthor{\bsnm{Miller}, \binits{S.A.}},
		\oauthor{\bsnm{Raithel}, \binits{G.}}:
		Using frequency detuning to improve the sensitivity of electric field
		measurements via electromagnetically induced transparency and autler-townes
		splitting in rydberg atoms.
		Applied Physics Letters
		\textbf{108}(17)
		(2016)
	\end{botherref}
	\endbibitem
	
	\bibitem[\protect\citeauthoryear{Kumar et~al.}{2017a}]{kumar2017atom}
	\begin{barticle}
		\bauthor{\bsnm{Kumar}, \binits{S.}},
		\bauthor{\bsnm{Fan}, \binits{H.}},
		\bauthor{\bsnm{K{\"u}bler}, \binits{H.}},
		\bauthor{\bsnm{Sheng}, \binits{J.}},
		\bauthor{\bsnm{Shaffer}, \binits{J.P.}}:
		\batitle{Atom-based sensing of weak radio frequency electric fields using
			homodyne readout}.
		\bjtitle{Scientific reports}
		\bvolume{7}(\bissue{1}),
		\bfpage{42981}
		(\byear{2017})
	\end{barticle}
	\endbibitem
	
	\bibitem[\protect\citeauthoryear{Kumar et~al.}{2017b}]{kumar2017rydberg}
	\begin{barticle}
		\bauthor{\bsnm{Kumar}, \binits{S.}},
		\bauthor{\bsnm{Fan}, \binits{H.}},
		\bauthor{\bsnm{K{\"u}bler}, \binits{H.}},
		\bauthor{\bsnm{Jahangiri}, \binits{A.J.}},
		\bauthor{\bsnm{Shaffer}, \binits{J.P.}}:
		\batitle{Rydberg-atom based radio-frequency electrometry using frequency
			modulation spectroscopy in room temperature vapor cells}.
		\bjtitle{Optics Express}
		\bvolume{25}(\bissue{8}),
		\bfpage{8625}--\blpage{8637}
		(\byear{2017})
	\end{barticle}
	\endbibitem
	
	\bibitem[\protect\citeauthoryear{Hao et~al.}{2023}]{hao2023microwave}
	\begin{botherref}
		\oauthor{\bsnm{Hao}, \binits{J.-H.}},
		\oauthor{\bsnm{Jia}, \binits{F.-D.}},
		\oauthor{\bsnm{Cui}, \binits{Y.}},
		\oauthor{\bsnm{Wang}, \binits{Y.-H.}},
		\oauthor{\bsnm{Zhou}, \binits{F.}},
		\oauthor{\bsnm{Liu}, \binits{X.-B.}},
		\oauthor{\bsnm{Zhang}, \binits{J.}},
		\oauthor{\bsnm{Xie}, \binits{F.}},
		\oauthor{\bsnm{Bai}, \binits{J.-H.}},
		\oauthor{\bsnm{You}, \binits{J.-Q.}}, et al.:
		Microwave electrometry with rydberg atoms in a vapor cell using microwave
		amplitude modulation.
		Chinese Physics B
		(2023)
	\end{botherref}
	\endbibitem
	
	\bibitem[\protect\citeauthoryear{Shaffer and
		K{\"u}bler}{2018}]{shaffer2018read}
	\begin{bchapter}
		\bauthor{\bsnm{Shaffer}, \binits{J.}},
		\bauthor{\bsnm{K{\"u}bler}, \binits{H.}}:
		\bctitle{A read-out enhancement for microwave electric field sensing with
			rydberg atoms}.
		In: \bbtitle{Quantum Technologies 2018},
		vol. \bseriesno{10674},
		pp. \bfpage{39}--\blpage{49}
		(\byear{2018}).
		\bcomment{SPIE}
	\end{bchapter}
	\endbibitem
	
	\bibitem[\protect\citeauthoryear{Jing et~al.}{2020}]{jing2020atomic}
	\begin{barticle}
		\bauthor{\bsnm{Jing}, \binits{M.}},
		\bauthor{\bsnm{Hu}, \binits{Y.}},
		\bauthor{\bsnm{Ma}, \binits{J.}},
		\bauthor{\bsnm{Zhang}, \binits{H.}},
		\bauthor{\bsnm{Zhang}, \binits{L.}},
		\bauthor{\bsnm{Xiao}, \binits{L.}},
		\bauthor{\bsnm{Jia}, \binits{S.}}:
		\batitle{Atomic superheterodyne receiver based on microwave-dressed rydberg
			spectroscopy}.
		\bjtitle{Nature Physics}
		\bvolume{16}(\bissue{9}),
		\bfpage{911}--\blpage{915}
		(\byear{2020})
	\end{barticle}
	\endbibitem
	
	\bibitem[\protect\citeauthoryear{Fan et~al.}{2016}]{fan2016dispersive}
	\begin{barticle}
		\bauthor{\bsnm{Fan}, \binits{H.}},
		\bauthor{\bsnm{Kumar}, \binits{S.}},
		\bauthor{\bsnm{K{\"u}bler}, \binits{H.}},
		\bauthor{\bsnm{Shaffer}, \binits{J.}}:
		\batitle{Dispersive radio frequency electrometry using rydberg atoms in a
			prism-shaped atomic vapor cell}.
		\bjtitle{Journal of Physics B: Atomic, Molecular and Optical Physics}
		\bvolume{49}(\bissue{10}),
		\bfpage{104004}
		(\byear{2016})
	\end{barticle}
	\endbibitem
	
	\bibitem[\protect\citeauthoryear{Yang et~al.}{2023}]{yang2023enhancing}
	\begin{barticle}
		\bauthor{\bsnm{Yang}, \binits{W.}},
		\bauthor{\bsnm{Jing}, \binits{M.}},
		\bauthor{\bsnm{Zhang}, \binits{H.}},
		\bauthor{\bsnm{Zhang}, \binits{L.}},
		\bauthor{\bsnm{Xiao}, \binits{L.}},
		\bauthor{\bsnm{Jia}, \binits{S.}}:
		\batitle{Enhancing the sensitivity of atom-based microwave-field electrometry
			using a mach-zehnder interferometer}.
		\bjtitle{Physical Review Applied}
		\bvolume{19}(\bissue{6}),
		\bfpage{064021}
		(\byear{2023})
	\end{barticle}
	\endbibitem
	
	\bibitem[\protect\citeauthoryear{Xiao et~al.}{1995}]{xiao1995measurement}
	\begin{barticle}
		\bauthor{\bsnm{Xiao}, \binits{M.}},
		\bauthor{\bsnm{Li}, \binits{Y.-q.}},
		\bauthor{\bsnm{Jin}, \binits{S.-z.}},
		\bauthor{\bsnm{Gea-Banacloche}, \binits{J.}}:
		\batitle{Measurement of dispersive properties of electromagnetically induced
			transparency in rubidium atoms}.
		\bjtitle{Physical Review Letters}
		\bvolume{74}(\bissue{5}),
		\bfpage{666}
		(\byear{1995})
	\end{barticle}
	\endbibitem
	
	\bibitem[\protect\citeauthoryear{Hau et~al.}{1999}]{hau1999light}
	\begin{barticle}
		\bauthor{\bsnm{Hau}, \binits{L.V.}},
		\bauthor{\bsnm{Harris}, \binits{S.E.}},
		\bauthor{\bsnm{Dutton}, \binits{Z.}},
		\bauthor{\bsnm{Behroozi}, \binits{C.H.}}:
		\batitle{Light speed reduction to 17 metres per second in an ultracold atomic
			gas}.
		\bjtitle{Nature}
		\bvolume{397}(\bissue{6720}),
		\bfpage{594}--\blpage{598}
		(\byear{1999})
	\end{barticle}
	\endbibitem
	
	\bibitem[\protect\citeauthoryear{Phillips et~al.}{2001}]{phillips2001storage}
	\begin{barticle}
		\bauthor{\bsnm{Phillips}, \binits{D.F.}},
		\bauthor{\bsnm{Fleischhauer}, \binits{A.}},
		\bauthor{\bsnm{Mair}, \binits{A.}},
		\bauthor{\bsnm{Walsworth}, \binits{R.L.}},
		\bauthor{\bsnm{Lukin}, \binits{M.D.}}:
		\batitle{Storage of light in atomic vapor}.
		\bjtitle{Physical review letters}
		\bvolume{86}(\bissue{5}),
		\bfpage{783}
		(\byear{2001})
	\end{barticle}
	\endbibitem
	
	\bibitem[\protect\citeauthoryear{Purves et~al.}{2004}]{purves2004refractive}
	\begin{barticle}
		\bauthor{\bsnm{Purves}, \binits{G.}},
		\bauthor{\bsnm{Jundt}, \binits{G.}},
		\bauthor{\bsnm{Adams}, \binits{C.}},
		\bauthor{\bsnm{Hughes}, \binits{I.}}:
		\batitle{Refractive index measurements by probe-beam deflection}.
		\bjtitle{The European Physical Journal D-Atomic, Molecular, Optical and Plasma
			Physics}
		\bvolume{29}(\bissue{3}),
		\bfpage{433}--\blpage{436}
		(\byear{2004})
	\end{barticle}
	\endbibitem
	
	\bibitem[\protect\citeauthoryear{Berweger et~al.}{2023}]{berweger2023closed}
	\begin{barticle}
		\bauthor{\bsnm{Berweger}, \binits{S.}},
		\bauthor{\bsnm{Artusio-Glimpse}, \binits{A.B.}},
		\bauthor{\bsnm{Rotunno}, \binits{A.P.}},
		\bauthor{\bsnm{Prajapati}, \binits{N.}},
		\bauthor{\bsnm{Christesen}, \binits{J.D.}},
		\bauthor{\bsnm{Moore}, \binits{K.R.}},
		\bauthor{\bsnm{Simons}, \binits{M.T.}},
		\bauthor{\bsnm{Holloway}, \binits{C.L.}}:
		\batitle{Closed-loop quantum interferometry for phase-resolved rydberg-atom
			field sensing}.
		\bjtitle{Physical Review Applied}
		\bvolume{20}(\bissue{5}),
		\bfpage{054009}
		(\byear{2023})
	\end{barticle}
	\endbibitem
	
	\bibitem[\protect\citeauthoryear{Aharonov et~al.}{1988}]{aharonov1988result}
	\begin{barticle}
		\bauthor{\bsnm{Aharonov}, \binits{Y.}},
		\bauthor{\bsnm{Albert}, \binits{D.Z.}},
		\bauthor{\bsnm{Vaidman}, \binits{L.}}:
		\batitle{How the result of a measurement of a component of the spin of a
			spin-1/2 particle can turn out to be 100}.
		\bjtitle{Physical review letters}
		\bvolume{60}(\bissue{14}),
		\bfpage{1351}
		(\byear{1988})
	\end{barticle}
	\endbibitem
	
	\bibitem[\protect\citeauthoryear{Ritchie et~al.}{1991}]{ritchie1991realization}
	\begin{barticle}
		\bauthor{\bsnm{Ritchie}, \binits{N.}},
		\bauthor{\bsnm{Story}, \binits{J.G.}},
		\bauthor{\bsnm{Hulet}, \binits{R.G.}}:
		\batitle{Realization of a measurement of a ‘‘weak value’’}.
		\bjtitle{Physical review letters}
		\bvolume{66}(\bissue{9}),
		\bfpage{1107}
		(\byear{1991})
	\end{barticle}
	\endbibitem
	
	\bibitem[\protect\citeauthoryear{Gorodetski et~al.}{2012}]{gorodetski2012weak}
	\begin{barticle}
		\bauthor{\bsnm{Gorodetski}, \binits{Y.}},
		\bauthor{\bsnm{Bliokh}, \binits{K.}},
		\bauthor{\bsnm{Stein}, \binits{B.}},
		\bauthor{\bsnm{Genet}, \binits{C.}},
		\bauthor{\bsnm{Shitrit}, \binits{N.}},
		\bauthor{\bsnm{Kleiner}, \binits{V.}},
		\bauthor{\bsnm{Hasman}, \binits{E.}},
		\bauthor{\bsnm{Ebbesen}, \binits{T.}}:
		\batitle{Weak measurements of light chirality with a plasmonic slit}.
		\bjtitle{Physical review letters}
		\bvolume{109}(\bissue{1}),
		\bfpage{013901}
		(\byear{2012})
	\end{barticle}
	\endbibitem
	
	\bibitem[\protect\citeauthoryear{Dixon et~al.}{2009}]{dixon2009ultrasensitive}
	\begin{barticle}
		\bauthor{\bsnm{Dixon}, \binits{P.B.}},
		\bauthor{\bsnm{Starling}, \binits{D.J.}},
		\bauthor{\bsnm{Jordan}, \binits{A.N.}},
		\bauthor{\bsnm{Howell}, \binits{J.C.}}:
		\batitle{Ultrasensitive beam deflection measurement via interferometric weak
			value amplification}.
		\bjtitle{Physical review letters}
		\bvolume{102}(\bissue{17}),
		\bfpage{173601}
		(\byear{2009})
	\end{barticle}
	\endbibitem
	
	\bibitem[\protect\citeauthoryear{Xu et~al.}{2013}]{xu2013phase}
	\begin{barticle}
		\bauthor{\bsnm{Xu}, \binits{X.-Y.}},
		\bauthor{\bsnm{Kedem}, \binits{Y.}},
		\bauthor{\bsnm{Sun}, \binits{K.}},
		\bauthor{\bsnm{Vaidman}, \binits{L.}},
		\bauthor{\bsnm{Li}, \binits{C.-F.}},
		\bauthor{\bsnm{Guo}, \binits{G.-C.}}:
		\batitle{Phase estimation with weak measurement using a white light source}.
		\bjtitle{Physical review letters}
		\bvolume{111}(\bissue{3}),
		\bfpage{033604}
		(\byear{2013})
	\end{barticle}
	\endbibitem
	
	\bibitem[\protect\citeauthoryear{Brunner and
		Simon}{2010}]{brunner2010measuring}
	\begin{barticle}
		\bauthor{\bsnm{Brunner}, \binits{N.}},
		\bauthor{\bsnm{Simon}, \binits{C.}}:
		\batitle{Measuring small longitudinal phase shifts: weak measurements or
			standard interferometry?}
		\bjtitle{Physical review letters}
		\bvolume{105}(\bissue{1}),
		\bfpage{010405}
		(\byear{2010})
	\end{barticle}
	\endbibitem
	
	\bibitem[\protect\citeauthoryear{Zhou et~al.}{2012}]{zhou2012experimental}
	\begin{barticle}
		\bauthor{\bsnm{Zhou}, \binits{X.}},
		\bauthor{\bsnm{Xiao}, \binits{Z.}},
		\bauthor{\bsnm{Luo}, \binits{H.}},
		\bauthor{\bsnm{Wen}, \binits{S.}}:
		\batitle{Experimental observation of the spin hall effect of light on a
			nanometal film via weak measurements}.
		\bjtitle{Physical Review A}
		\bvolume{85}(\bissue{4}),
		\bfpage{043809}
		(\byear{2012})
	\end{barticle}
	\endbibitem
	
	\bibitem[\protect\citeauthoryear{Viza et~al.}{2013}]{viza2013weak}
	\begin{barticle}
		\bauthor{\bsnm{Viza}, \binits{G.I.}},
		\bauthor{\bsnm{Mart{\'\i}nez-Rinc{\'o}n}, \binits{J.}},
		\bauthor{\bsnm{Howland}, \binits{G.A.}},
		\bauthor{\bsnm{Frostig}, \binits{H.}},
		\bauthor{\bsnm{Shomroni}, \binits{I.}},
		\bauthor{\bsnm{Dayan}, \binits{B.}},
		\bauthor{\bsnm{Howell}, \binits{J.C.}}:
		\batitle{Weak-values technique for velocity measurements}.
		\bjtitle{Optics letters}
		\bvolume{38}(\bissue{16}),
		\bfpage{2949}--\blpage{2952}
		(\byear{2013})
	\end{barticle}
	\endbibitem
	
	\bibitem[\protect\citeauthoryear{Starling et~al.}{2010}]{starling2010precision}
	\begin{barticle}
		\bauthor{\bsnm{Starling}, \binits{D.J.}},
		\bauthor{\bsnm{Dixon}, \binits{P.B.}},
		\bauthor{\bsnm{Jordan}, \binits{A.N.}},
		\bauthor{\bsnm{Howell}, \binits{J.C.}}:
		\batitle{Precision frequency measurements with interferometric weak values}.
		\bjtitle{Physical Review A}
		\bvolume{82}(\bissue{6}),
		\bfpage{063822}
		(\byear{2010})
	\end{barticle}
	\endbibitem
	
	\bibitem[\protect\citeauthoryear{Egan and Stone}{2012}]{egan2012weak}
	\begin{barticle}
		\bauthor{\bsnm{Egan}, \binits{P.}},
		\bauthor{\bsnm{Stone}, \binits{J.A.}}:
		\batitle{Weak-value thermostat with 0.2 mk precision}.
		\bjtitle{Optics letters}
		\bvolume{37}(\bissue{23}),
		\bfpage{4991}--\blpage{4993}
		(\byear{2012})
	\end{barticle}
	\endbibitem
	
	\bibitem[\protect\citeauthoryear{Salazar-Serrano
		et~al.}{2015}]{salazar2015enhancement}
	\begin{barticle}
		\bauthor{\bsnm{Salazar-Serrano}, \binits{L.}},
		\bauthor{\bsnm{Barrera}, \binits{D.}},
		\bauthor{\bsnm{Amaya}, \binits{W.}},
		\bauthor{\bsnm{Sales}, \binits{S.}},
		\bauthor{\bsnm{Pruneri}, \binits{V.}},
		\bauthor{\bsnm{Capmany}, \binits{J.}},
		\bauthor{\bsnm{Torres}, \binits{J.}}:
		\batitle{Enhancement of the sensitivity of a temperature sensor based on fiber
			bragg gratings via weak value amplification}.
		\bjtitle{Optics letters}
		\bvolume{40}(\bissue{17}),
		\bfpage{3962}--\blpage{3965}
		(\byear{2015})
	\end{barticle}
	\endbibitem
	
	\bibitem[\protect\citeauthoryear{Zhang et~al.}{2015}]{PhysRevLett.114.210801}
	\begin{barticle}
		\bauthor{\bsnm{Zhang}, \binits{L.}},
		\bauthor{\bsnm{Datta}, \binits{A.}},
		\bauthor{\bsnm{Walmsley}, \binits{I.A.}}:
		\batitle{Precision metrology using weak measurements}.
		\bjtitle{Phys. Rev. Lett.}
		\bvolume{114},
		\bfpage{210801}
		(\byear{2015})
		\doiurl{10.1103/PhysRevLett.114.210801}
	\end{barticle}
	\endbibitem
	
	\bibitem[\protect\citeauthoryear{Xu et~al.}{2020}]{xu2020approaching}
	\begin{barticle}
		\bauthor{\bsnm{Xu}, \binits{L.}},
		\bauthor{\bsnm{Liu}, \binits{Z.}},
		\bauthor{\bsnm{Datta}, \binits{A.}},
		\bauthor{\bsnm{Knee}, \binits{G.C.}},
		\bauthor{\bsnm{Lundeen}, \binits{J.S.}},
		\bauthor{\bsnm{Lu}, \binits{Y.-q.}},
		\bauthor{\bsnm{Zhang}, \binits{L.}}:
		\batitle{Approaching quantum-limited metrology with imperfect detectors by
			using weak-value amplification}.
		\bjtitle{Physical Review Letters}
		\bvolume{125}(\bissue{8}),
		\bfpage{080501}
		(\byear{2020})
	\end{barticle}
	\endbibitem
	
	\bibitem[\protect\citeauthoryear{Qiu et~al.}{2014}]{qiu2014determination}
	\begin{botherref}
		\oauthor{\bsnm{Qiu}, \binits{X.}},
		\oauthor{\bsnm{Zhou}, \binits{X.}},
		\oauthor{\bsnm{Hu}, \binits{D.}},
		\oauthor{\bsnm{Du}, \binits{J.}},
		\oauthor{\bsnm{Gao}, \binits{F.}},
		\oauthor{\bsnm{Zhang}, \binits{Z.}},
		\oauthor{\bsnm{Luo}, \binits{H.}}:
		Determination of magneto-optical constant of fe films with weak measurements.
		Applied Physics Letters
		\textbf{105}(13)
		(2014)
	\end{botherref}
	\endbibitem
	
	\bibitem[\protect\citeauthoryear{Zhang et~al.}{2016}]{zhang2016optical}
	\begin{barticle}
		\bauthor{\bsnm{Zhang}, \binits{Y.}},
		\bauthor{\bsnm{Li}, \binits{D.}},
		\bauthor{\bsnm{He}, \binits{Y.}},
		\bauthor{\bsnm{Shen}, \binits{Z.}},
		\bauthor{\bsnm{He}, \binits{Q.}}:
		\batitle{Optical weak measurement system with common path implementation for
			label-free biomolecule sensing}.
		\bjtitle{Optics letters}
		\bvolume{41}(\bissue{22}),
		\bfpage{5409}--\blpage{5412}
		(\byear{2016})
	\end{barticle}
	\endbibitem
	
	\bibitem[\protect\citeauthoryear{Lyons et~al.}{2018}]{lyons2018noise}
	\begin{barticle}
		\bauthor{\bsnm{Lyons}, \binits{K.}},
		\bauthor{\bsnm{Howell}, \binits{J.C.}},
		\bauthor{\bsnm{Jordan}, \binits{A.N.}}:
		\batitle{Noise suppression in inverse weak value-based phase detection}.
		\bjtitle{Quantum Studies: Mathematics and Foundations}
		\bvolume{5},
		\bfpage{579}--\blpage{588}
		(\byear{2018})
	\end{barticle}
	\endbibitem
	
	\bibitem[\protect\citeauthoryear{Viza et~al.}{2016}]{viza2016complementary}
	\begin{barticle}
		\bauthor{\bsnm{Viza}, \binits{G.I.}},
		\bauthor{\bsnm{Mart{\'\i}nez-Rinc{\'o}n}, \binits{J.}},
		\bauthor{\bsnm{Liu}, \binits{W.-T.}},
		\bauthor{\bsnm{Howell}, \binits{J.C.}}:
		\batitle{Complementary weak-value amplification with concatenated
			postselections}.
		\bjtitle{Physical Review A}
		\bvolume{94}(\bissue{4}),
		\bfpage{043825}
		(\byear{2016})
	\end{barticle}
	\endbibitem
	
	\bibitem[\protect\citeauthoryear{Starling
		et~al.}{2010}]{starling2010continuous}
	\begin{barticle}
		\bauthor{\bsnm{Starling}, \binits{D.J.}},
		\bauthor{\bsnm{Dixon}, \binits{P.B.}},
		\bauthor{\bsnm{Williams}, \binits{N.S.}},
		\bauthor{\bsnm{Jordan}, \binits{A.N.}},
		\bauthor{\bsnm{Howell}, \binits{J.C.}}:
		\batitle{Continuous phase amplification with a sagnac interferometer}.
		\bjtitle{Physical Review A}
		\bvolume{82}(\bissue{1}),
		\bfpage{011802}
		(\byear{2010})
	\end{barticle}
	\endbibitem
	
	\bibitem[\protect\citeauthoryear{Song et~al.}{2021}]{song2021enhanced}
	\begin{barticle}
		\bauthor{\bsnm{Song}, \binits{M.}},
		\bauthor{\bsnm{Steinmetz}, \binits{J.}},
		\bauthor{\bsnm{Zhang}, \binits{Y.}},
		\bauthor{\bsnm{Nauriyal}, \binits{J.}},
		\bauthor{\bsnm{Lyons}, \binits{K.}},
		\bauthor{\bsnm{Jordan}, \binits{A.N.}},
		\bauthor{\bsnm{Cardenas}, \binits{J.}}:
		\batitle{Enhanced on-chip phase measurement by inverse weak value
			amplification}.
		\bjtitle{Nature Communications}
		\bvolume{12}(\bissue{1}),
		\bfpage{6247}
		(\byear{2021})
	\end{barticle}
	\endbibitem
	
	\bibitem[\protect\citeauthoryear{Mi{\v{c}}uda et~al.}{2014}]{mivcuda2014highly}
	\begin{botherref}
		\oauthor{\bsnm{Mi{\v{c}}uda}, \binits{M.}},
		\oauthor{\bsnm{Dol{\'a}kov{\'a}}, \binits{E.}},
		\oauthor{\bsnm{Straka}, \binits{I.}},
		\oauthor{\bsnm{Mikov{\'a}}, \binits{M.}},
		\oauthor{\bsnm{Du{\v{s}}ek}, \binits{M.}},
		\oauthor{\bsnm{Fiur{\'a}{\v{s}}ek}, \binits{J.}},
		\oauthor{\bsnm{Je{\v{z}}ek}, \binits{M.}}:
		Highly stable polarization independent mach-zehnder interferometer.
		Review of Scientific Instruments
		\textbf{85}(8)
		(2014)
	\end{botherref}
	\endbibitem
	
	\bibitem[\protect\citeauthoryear{Jordan and Siddiqi}{2024}]{jordan2024quantum}
	\begin{bbook}
		\bauthor{\bsnm{Jordan}, \binits{A.N.}},
		\bauthor{\bsnm{Siddiqi}, \binits{I.A.}}:
		\bbtitle{Quantum Measurement: Theory and Practice}.
		\bpublisher{Cambridge University Press}, \blocation{???}
		(\byear{2024})
	\end{bbook}
	\endbibitem
	
	\bibitem[\protect\citeauthoryear{Jordan et~al.}{2015}]{jordan2015heisenberg}
	\begin{barticle}
		\bauthor{\bsnm{Jordan}, \binits{A.N.}},
		\bauthor{\bsnm{Tollaksen}, \binits{J.}},
		\bauthor{\bsnm{Troupe}, \binits{J.E.}},
		\bauthor{\bsnm{Dressel}, \binits{J.}},
		\bauthor{\bsnm{Aharonov}, \binits{Y.}}:
		\batitle{Heisenberg scaling with weak measurement: a quantum state
			discrimination point of view}.
		\bjtitle{Quantum Studies: Mathematics and Foundations}
		\bvolume{2},
		\bfpage{5}--\blpage{15}
		(\byear{2015})
	\end{barticle}
	\endbibitem
	
	\bibitem[\protect\citeauthoryear{Kofman
		et~al.}{2012}]{kofman2012nonperturbative}
	\begin{barticle}
		\bauthor{\bsnm{Kofman}, \binits{A.G.}},
		\bauthor{\bsnm{Ashhab}, \binits{S.}},
		\bauthor{\bsnm{Nori}, \binits{F.}}:
		\batitle{Nonperturbative theory of weak pre-and post-selected measurements}.
		\bjtitle{Physics Reports}
		\bvolume{520}(\bissue{2}),
		\bfpage{43}--\blpage{133}
		(\byear{2012})
	\end{barticle}
	\endbibitem
	
	\bibitem[\protect\citeauthoryear{Yin et~al.}{2021}]{yin2021improving}
	\begin{barticle}
		\bauthor{\bsnm{Yin}, \binits{P.}},
		\bauthor{\bsnm{Zhang}, \binits{W.-H.}},
		\bauthor{\bsnm{Xu}, \binits{L.}},
		\bauthor{\bsnm{Liu}, \binits{Z.-G.}},
		\bauthor{\bsnm{Zhuang}, \binits{W.-F.}},
		\bauthor{\bsnm{Chen}, \binits{L.}},
		\bauthor{\bsnm{Gong}, \binits{M.}},
		\bauthor{\bsnm{Ma}, \binits{Y.}},
		\bauthor{\bsnm{Peng}, \binits{X.-X.}},
		\bauthor{\bsnm{Li}, \binits{G.-C.}}, \betal:
		\batitle{Improving the precision of optical metrology by detecting fewer
			photons with biased weak measurement}.
		\bjtitle{Light: Science \& Applications}
		\bvolume{10}(\bissue{1}),
		\bfpage{103}
		(\byear{2021})
	\end{barticle}
	\endbibitem
	
	\bibitem[\protect\citeauthoryear{Fan et~al.}{2015}]{fan2015atom}
	\begin{barticle}
		\bauthor{\bsnm{Fan}, \binits{H.}},
		\bauthor{\bsnm{Kumar}, \binits{S.}},
		\bauthor{\bsnm{Sedlacek}, \binits{J.}},
		\bauthor{\bsnm{K{\"u}bler}, \binits{H.}},
		\bauthor{\bsnm{Karimkashi}, \binits{S.}},
		\bauthor{\bsnm{Shaffer}, \binits{J.P.}}:
		\batitle{Atom based rf electric field sensing}.
		\bjtitle{Journal of Physics B: Atomic, Molecular and Optical Physics}
		\bvolume{48}(\bissue{20}),
		\bfpage{202001}
		(\byear{2015})
	\end{barticle}
	\endbibitem
	
	\bibitem[\protect\citeauthoryear{Caves}{1981}]{PhysRevD.23.1693}
	\begin{barticle}
		\bauthor{\bsnm{Caves}, \binits{C.M.}}:
		\batitle{Quantum-mechanical noise in an interferometer}.
		\bjtitle{Phys. Rev. D}
		\bvolume{23},
		\bfpage{1693}--\blpage{1708}
		(\byear{1981})
		\doiurl{10.1103/PhysRevD.23.1693}
	\end{barticle}
	\endbibitem
	
	
		\bibitem[\protect\citeauthoryear{Jiang et~al.}{2024}]{Jiang:24}
	\begin{barticle}
		\bauthor{\bsnm{Jiang}, \binits{Y.}},
		\bauthor{\bsnm{Wu}, \binits{J.}},
		\bauthor{\bsnm{Ge}, \binits{R.}},
		\bauthor{\bsnm{Zhang}, \binits{Z.}}:
		\batitle{Observation of the spin hall effect of light by a single-photon
			detector}.
		\bjtitle{Opt. Lett.}
		\bvolume{49}(\bissue{11}),
		\bfpage{3014}--\blpage{3017}
		(\byear{2024})
		\doiurl{10.1364/OL.522132}
	\end{barticle}
	\endbibitem
	
		\bibitem[\protect\citeauthoryear{Jiang et~al.}{2025}]{Data}
	\begin{barticle}
		\bauthor{\bsnm{Jiang}, \binits{Y.}},
		\bauthor{\bsnm{Wu}, \binits{J.}},
		\bauthor{\bsnm{Shi}, \binits{M.}},
		\bauthor{\bsnm{Xiao}, \binits{Z.}},
		\bauthor{\bsnm{Guo}, \binits{F.}},
		\bauthor{\bsnm{Zheng}, \binits{H.}},
		\bauthor{\bsnm{Zhang}, \binits{Z.}},:
		\batitle{Data for 'Quantum Weak Measurement Amplifies Dispersion Signal of Rydberg Atomic System'}.
		\bjtitle{figshare},
		(\byear{2025})
		\doiurl{10.6084/m9.figshare.28292981.v1  }
	\end{barticle}
	\endbibitem
	
\end{thebibliography}

\section*{Acknowledgements}
 The authors are grateful for the support from China National Natural Science Foundation
under contract No. 12335002, 12375078, 11975028. Also supported by "the Fundamental Research Funds for the Central Universities".

\section*{Author contribution}
Y.J. and J.W. proposed idea and completed the experiment. M.S., H.Z. and Z.Z. provided experimental conditions and gave guidance. Y.J. completed theoretical details, data analysis and manuscript. Discussions with Z.X. and F.G. helped complete the draft.

\section*{Competing interests}
The authors declare no competing interests.
 
\end{document}